\documentclass[twocolumn]{article}
\usepackage{amsmath}
\usepackage{epsfig}
\usepackage{graphicx}
\usepackage{amsfonts}
\usepackage{amssymb}
\usepackage{epsf}

\begin{document}

\author{Rafael~Bautista-Mena\\CEIBA Center, Universidad de los Andes, Bogot\'{a}, Colombia}
\title{Alice and Bob and Hendrik}
\maketitle

\begin{abstract}
This paper offers an alternative approach to discussing both the
principle of relativity and the derivation of the Lorentz
transformations. This approach uses the idea that there may not be
a preferred inertial frame through a privileged access to
information about events. In classroom discussions, it has been my
experience that this approach produces some lively arguments.
\end{abstract}

\section{Concepts and conventions}
Suppose that two inertial ``observers'', from now on named Alice
and Bob, each attached to a reference frame in one spatial
dimension and time, need to exchange information about an event
$E$ that is a part of an information set to which both have access
to. In discussions about kinematics, the information set about $E$
reduces to three items: 1.- The fact that it took place (for
instance, as registered by a ``click'' in standard detectors that
both Alice and Bob are endowed with.) 2.- The spatial coordinate
for $E$, and 3.- the associated time. In such discussions, the
first item is almost always taken for granted. For the purposes of
this work, the only proviso I will make is to call ``events'' only
those described in item 1, and avoid the more widespread usage,
where ``event'' tends to be more or less automatically identified
with a space-time point. Items 2 and 3 are the usual concern of
kinematics, and as is well known, the communication of these data
between Alice and Bob goes via a set of simple, linear
transformation equations, namely the Lorentz transformations.
\newline Another element necessary for the discussion that follows is
to assume the existence of some standard ``messenger'', which may
be produced by any event, and is by definition the
fastest\footnote{As measured in each of their frames.} entity
known to the observers to be apt to carry encoded messages across
empty space. As an integral part of the definition of their
frames, Alice and Bob are capable of either encoding or decoding
messages, using some universal code. Such messages always carry
the information about those events registered in their frames. By
assumption, any event that Alice is capable of registering with
her detector will also be detectable by Bob.
\newline Bob's frame of reference will be drawn as a Cartesian
frame, with the vertical time axis perpendicular to the horizontal
space axis. By definition, the origin corresponds to the event of
coincidence with that of Alice's frame. This event is labelled $O$
in Figure 1.
\newline Alice's choice of axes for encoding her information about
events is completely defined by the angles (see Figure 1) $\alpha$
and $\gamma$ that her time and space axes respectively make with
those of Bob's frame. I adopt the convention that, as drawn, these
angles are positive, and satisfy the constraint
\begin{equation}
\alpha+\gamma\leq \frac{\pi}{2}
\end{equation}

\begin{figure}[t] \label{Fi:fig}
\begin{center}
\epsfig{file=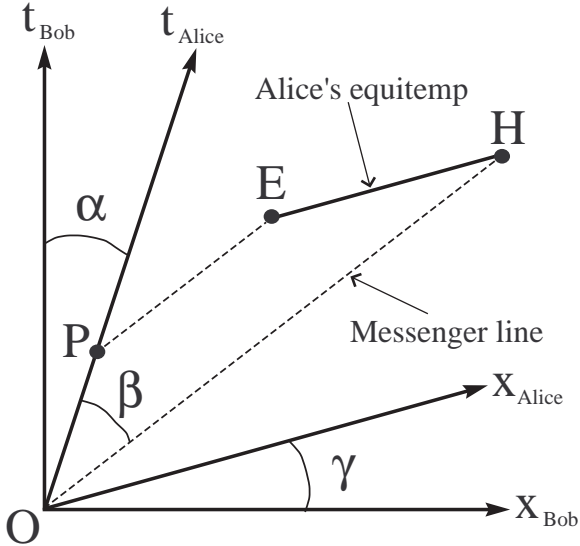, clip= ,width=1.0\linewidth }
\caption{Angle conventions between Alice's and Bob's frames}
\end{center}
\end{figure}

Aside from this constraint, Alice's choice of frame would in
principle be arbitrary.
\newline Measured from Alice's time
axis, the world lines followed by the messenger in her frame tilt
an angle $\beta$. All lines parallel to Alice's $x$ axis will be
called ``equitemps'', and all lines parallel to her time axis will
be called ``equilocs", after the denomination used in
Mermin~\cite{mermin1}.\newline It is important to emphasize that
the particular choice of perpendicular axes for Bob's frame is
made only for ease of exposition. In fact, all that matters is the
relative angle between Bob's and Alice's corresponding axes. In
what follows, methods of Euclidean geometry will be used within a
context that is usually associated with Minkowski's space-time.
Further details as to why and how this may be done can be found in
Brill and Jacobson~\cite{brilljacob}.
\subsection{Locality of the observers}
The act of interpreting, or decoding, a message is, by assumption,
local in character. I will model this assumption by locating Alice
at some definite, fixed point. By convention, Alice will be
located at the spatial origin for all time. As a consequence,
Alice will learn about the occurrence of event $E$, that took
place say at frame coordinates $(x_{E},t_{E})$ only at a later
time $t_{E}+x_{E}/c$, where $c$ is the speed of the standard
messenger in her frame. In a symmetric fashion, if Alice wanted to
be causally connected with event $E$, then the latest moment at
which she could send a ``triggering'' message would be
$t_{E}-x_{E}/c$. This particular event, Alice's delivery of the
latest signal that could connect her to $E$, is shown at the
vertex $P$ in Figure 1. As an obvious extension, Alice could be
causally connected to any event that may be triggered along the
world-line connecting Alice with event $E$. Another event of
interest is the ``horizon'' event $H$. This is the event
simultaneous with $E$, triggered by a messenger sent out from $O$
along the positive $x$ axis. Notice that this is the farthest
point to which Alice may expect to be causally connected with
before or simultaneously with event $E$.

\subsection{Accessible sets}
An accessible event, conditional on events $E$ and $O$, is one for
which Alice may be able to have a causal connection with before or
at most simultaneously with the occurrence of event $E$. The
accessible set $A$, conditional on events $O$ and $E$, is the set
of all space-time points with which Alice could establish a causal
connection, right after her time $t=0$ and up until the frame time
at which $E$ takes place. From this description, it is clear that
$E$ must lie within the ``causal cone'' defined by the messenger.
In the space-time diagram shown in Figure 1, the polygon $OPEHO$
corresponds to the accessible set conditional on events $O$ and
$E$.\newline The intuition for $A$ is to think of it as composed
by member ``sites'', all identical in their properties, and
distinguished only by their space-time coordinates. Each member of
$A$ is equally capable of hosting a single event.\footnote{Here
the ``site'' of an event is assumed to correspond to a single
member of the set. In principle, one could consider the
possibility of a larger subset of $A$ as the site of a single
event, in which case the individual space-time coordinates of the
elements in the subset would fail to provide any meaningful
information about the event. But this case looks more like quantum
physics. The present discussion is fully contained within a
classical context.} Therefore, $A$ describes Alice's capacity to
influence events along the positive $x$ axis between $t=0$ and the
time corresponding to event $E$. Alternatively, $A$ may also be
seen as the set containing the maximum amount of information
(potential events) generated between those times, that Alice may
expect to collect.

\section{The relativity principle}
Given the same constraints: a common event $O$, and an external,
independent event $E$, no inertial observer may expect to be
causally connected to more sites, or to be able to have access to
more information than any other. In more mundane terms, given the
same prior information, Alice may not know anything that Bob
wouldn't know too, nor vice versa.
\newline One way to make
operative this form of the relativity principle is to assign a
measure $I(A)$ to set $A$. I will make the following assumption:\\
\newline \emph{ The Euclidean area of the accessible set bounded by the polygon $OPEHO$ is a
direct measure of the maximum number of events about which Alice
may have knowledge, conditional on events $O$ and $E$.}\footnote{I
am aware that this implies attaching to the set $A$, and to
space-time in general, a topology different from the usually
assumed for continuous space-time. In fact, it would have to be
based on finite, or at most, countable sets. But the size of the
corresponding ``space-time cells'' could be made as small as
desired, as long as they were finite.} \\
\newline Without providing a proof, it seems reasonable to suppose that
this statement is fully consistent with the properties of
homogeneity and isotropy of flat space-time.

\section{Derivation of results}
Let's begin by computing the coordinates for all four events
defining Alice's accessible set, as determined by Bob. Figure 1
shows the polygon $OPEHO$ and the associated angles.
\newline In self-evident notation, the coordinates for each vertex
as functions of Bob's coordinates $(X,T)$ for event $E$ are given
by:
\begin{equation}
x_{P}=\frac{\sin\alpha}{\sin\beta}\Big[T\sin(\alpha+\beta)-X\cos(\alpha+\beta)\Big],
\end{equation}
\begin{equation}
t_{P}=x_{P}\cot\alpha,
\end{equation}
\begin{equation}
x_{H}=\frac{T-X\tan\gamma}{\cot(\alpha+\beta)-\tan\gamma},
\end{equation}
\begin{equation}
t_{H}=x_{H}\cot(\alpha+\beta).
\end{equation}
The measure of the accessible set, corresponding to the area
bounded by $OPEHO$ is
\begin{equation}\label{E:measure}
I(A)=\frac{1}{2}\Big[Xt_{P}-Tx_{P}+x_{H}T-t_{H}X\Big].
\end{equation}
Equation~(\ref{E:measure}) may be rewritten as follows:
\begin{equation}\label{E:measure2}
2I(A)=h_{1}T^{2}+h_{2}X^{2}+h_{3}XT.
\end{equation}
The principle of relativity, as stated here, now requires that
this measure be frame invariant. In other words, the $h_{i}$'s
in~(\ref{E:measure2}) ought to be universal constants. It is not
meant here that those coefficients are new physical constants, in
the sense that Planck's constant or the charge of the electron
are. But rather, that the corresponding algebraic expressions for
the $h_{i}$'s must reduce, in a trivial way, to simple numerical
values. Therefore, as an immediate consequence of the relativity
principle, there follows:
\begin{equation}\label{E:relp}
h_{i}=constant.
\end{equation}
Their explicit forms are the following:
\begin{multline}\label{E:h1}
h_{1}=-\frac{\sin\alpha}{\sin\beta}\sin(\alpha+\beta)\\+\frac{1}{\cot(\alpha+\beta)-\tan\gamma},
\end{multline}
\begin{multline}\label{E:h2}
h_{2}=-\frac{\cos\alpha}{\sin\beta}\cos(\alpha+\beta)\\+\frac{\cot(\alpha+\beta)\tan\gamma}{\cot(\alpha+\beta)-\tan\gamma},
\end{multline}
\begin{multline}\label{E:h3}
h_{3}=\frac{\cos\alpha \sin(\alpha+\beta)+ \sin\alpha
\cos(\alpha+\beta)}{\sin\beta} \\-\frac{\cot(\alpha+\beta)+
\tan\gamma}{\cot(\alpha+\beta)- \tan\gamma}.
\end{multline}
These rather lengthy expressions may be more easily handled using
the following shorthand notation: $v\equiv\tan\alpha \mbox{;
}w\equiv\tan(\alpha+\beta) \mbox{; }z\equiv\tan\gamma$.
Now~(\ref{E:h1}),~(\ref{E:h2}) and~(\ref{E:h3}) look as follows:
\begin{gather}
h_{1}=-\frac{wv}{w-v}+\frac{w}{1-zw},\label{E:hh1}\\
h_{2}=-\frac{1}{w-v}+\frac{z}{1-zw},\label{E:hh2}\\
h_{3}=\frac{w+v}{w-v}-\frac{1+zw}{1-zw}.\label{E:hh3}
\end{gather}
Inspection of~(\ref{E:hh1}),~(\ref{E:hh2}) and~(\ref{E:hh3}) leads
to the following identity:
\begin{equation}\label{E:constraint}
\frac{h_{1}}{w}+h_{2}w+h_{3}=0.
\end{equation}
From the statement of the relativity principle in~(\ref{E:relp}),
equation~(\ref{E:constraint}) implies that $w$ is equal to some
constant value. Therefore:
\begin{equation}\label{E:slope}
\alpha+\beta=constant.
\end{equation}
Then, irrespective of the choice of frame, the relative slope
associated with the speed of the messenger is fixed.
\newline Equation~(\ref{E:hh1}) can be rearranged as follows:
\begin{multline}\label{E:frame}
vz(w^{2}-h_{1}w)+w^{2}h_{1}z\\+(h_{1}-2w)v+w^{2}-h_{1}w=0.
\end{multline}
Since both $v$ and $z$ represent trigonometric functions, and
since~(\ref{E:frame}) must be an identity, quadratic terms should
be linearly independent from linear terms, therefore the
coefficient in the quadratic term must vanish, leaving as its only
feasible solution:
\begin{equation}\label{E:solw}
w=h_{1}.
\end{equation}
Notice that $w=0$ is not a feasible solution, for it doesn't
solve~(\ref{E:constraint}). With this result,
equation~(\ref{E:frame}) reduces to:
\begin{equation}\label{E:zv}
h^{2}_{1}z-v=0.
\end{equation}
From this relation follows that, if the principle as stated
by~(\ref{E:relp}) is to be upheld, then the choice of axes by
Alice is constrained by~(\ref{E:zv}). This relationship is, by the
way, the best justification of why $h_{1}$ must be different from
zero, for otherwise, Alice wouldn't have a choice at all, or put
another way, it would deny the existence of any reference frame.
\newline Using~(\ref{E:solw}), equation~(\ref{E:constraint})
becomes:
\begin{equation}\label{E:constraint2}
1+h_{1}h_{2}+h_{3}=0.
\end{equation}
Substitution of~(\ref{E:solw}) and~(\ref{E:zv}) into~(\ref{E:hh2})
yields:
\begin{equation}\label{E:h1h2}
h_{1}h_{2}=-1.
\end{equation}
This last result, combined with~(\ref{E:constraint2}) produce:
\begin{equation}\label{E:solh3}
h_{3}=0.
\end{equation}
These findings for the $h_{i}$'s lead back to~(\ref{E:measure2}),
which now reduces to:
\begin{equation}\label{E:interval}
2I(A)=h_{1}X^{2}-\frac{T^{2}}{h_{1}}.
\end{equation}
In this expression, it is always possible to set $h_{1}=1$,
because this is just a rescaling of the ruler and the ``tick'' of
the clock used by Alice. Then~(\ref{E:interval}) is easily
recognizable as the Minkowski square of the space-time interval.
This same choice makes $w=1$, which, going back
to~(\ref{E:slope}), produces the neat result:
\begin{equation}\label{E:diag}
\alpha+\beta=\frac{\pi}{4}.
\end{equation}
Then, the messenger's slope must cut in halves the quadrant of
Bob's frame. Finally,~(\ref{E:zv}) simplifies to:
\begin{equation}\label{E:lorentz}
v=z.
\end{equation}
Using the convention established earlier for $\alpha$ and
$\gamma$, the last equation is equivalent to say that they are
equal. Therefore, Alice's axes are also placed symmetrically
around the line of the messenger. This geometrical arrangement is
well known: Bob's and Alice's frames are connected by the Lorentz
transformation.
\newline The second consequence that follows
from~(\ref{E:lorentz}) is that it makes obvious that in Alice's
frame the speed of the messenger is the same as in Bob's frame.
Therefore, there exists one messenger whose speed is the same in
all frames of reference.
\section{Discussion}
In the present work I have derived both the necessity of the
existence of a messenger with an invariant speed in all frames of
reference and the Lorentz transformations, starting from the
principle of relativity, stated as a symmetry in the access to
causal connections. In more relaxed terms, this approach
establishes the impossibility to tell the state of inertial motion
via the access to different ``amounts of information'' between
reference frames. This approach
relies on two assumptions:\\
\newline 1. The local character of any encoding/decoding
capable ``observer''.
\newline 2. The measure $I(A)$ as the correct invariant quantity.\\
\newline Traditionally, the question `Why the
Lorentz transformation?' has been answered with `Because it is the
only solution consistent with the relativity principle.' The
present work instead addresses the question `Which way to the
relativity principle?' Within the context of this paper, the
principle has been spelled out through the invariance of the
measure of the conditional accessible set $I(A)$.
\newline A question raised by this approach may be why
it works. It is not new to obtain the Lorentz transformations from
the relativity principle, plus additional assumptions about the
properties of flat space-time, as it has been shown in several
excellent articles (see, for instance, L\'{e}vy-Leblond
\cite{levy}, Mermin \cite{mermin2}, Lee and Kalotas
\cite{leekalotas}.) The only difference in my approach is the
expression of the principle in terms of a kind of information
democracy, which is closer in spirit to the intend in Field
\cite{field}, who arrives at the Lorentz transformations from a
postulated space-time exchange invariance. To see the connection
with other treatments, recall that a universal messenger generates
a causal ordering on the future cone. Therefore, the relativity
principle imposes a causal structure on set $A$. Turning this
argument around, suppose now that we would want to have a set $A$
with a postulated causal structure.\footnote{See an interesting
approach to the Lorentz transformations from this angle in
\cite{zeeman}.} Suppose also that Alice triggered an event timed
between events $O$ and $E$. Since Alice and Bob are equivalent in
their capacity to register events, Bob would learn about such
event. But he could not register this event as having occurred
either before $O$ or after $E$, because that would violate the
assumption of a causal structure for $A$. Therefore, all the
events that Alice would trigger between $O$ and $E$, are the same
ones that Bob could detect too, no more and no less. This argument
sheds light on why the invariance of the measure $I(A)$ acts as a
substitute for the conventional statement of the relativity
principle. Then, the main contribution of this particular
formulation is its approach to relativity from an event-counting
concept, represented (as proxy) by a Euclidean measure. On the
other hand, one limitation is that it starts from the assumption
that the correct transformation relation between inertial frames
is linear.
\newline This approach is open to criticism, among other reasons, on
the basis that it looks like a step back toward anthropocentrism,
through my recourse to terms such as ``information'',
``encoding'', ``decoding'', and others of a similar nature. I
always bear in mind the now famous retort `Whose information?'
Nevertheless, I believe that my use of such terms only highlights
the limitations of language. After all, in the case of, say, an
elastic collision between two electrons, we use terms such as
``interaction'' to refer to the exchange of momentum between the
particles, only out of well established tradition.
\section{Acknowledgements}
I wish to thank Ana Rey, Juan Restrepo, Alonso Botero and Jorge
Villalobos for their useful comments and their kind help.

\end{document}